\documentstyle[11pt,aaspp,epsf]{article}

\begin{document}

\title{A Second Bright Source Detected Near SN1987A}

\author{Peter Nisenson} 

\affil{Harvard-Smithsonian Center for Astrophysics \\
60 Garden Street, Cambridge, MA 02138 \\
e-mail: pnisenson@cfa.harvard.edu}

\author {Costas Papaliolios}

\affil{Harvard-Smithsonian Center for Astrophysics \\
60 Garden Street, Cambridge, MA 02138 \\
e-mail: cpapaliolios@cfa.harvard.edu}

\newpage

\begin{abstract}

Speckle interferometry observations, made just 30 and 38 days after the
explosion of supernova SN1987A (SN) (which was first seen on February 23,
1987), showed evidence for a bright source, separated from the SN by only 60
mas.  Reprocessing of that data using new image reconstruction algorithms has
resulted in much cleaner images which not only clearly show the bright spot
reported in 1987, but also a 2nd spot on the opposite side of the SN with a
larger spatial separation. If the spots were ejected from the SN then the
velocities of the spots are relativistic and the 2nd spot appears to
be superluminal and must be blue-shifted. We explore the consequences
of these results on the geometry of the SN1987A system, and we 
conclude that our observations may well be evidence for a relativistic jet
emanating from the supernova.

\end{abstract}

Subject Headings: stars: supernovae: individual (SN1987A);
techniques:interferometric; gamma rays: bursts

\section{Introduction}

Speckle interferometry observations, made just 30 and 38 days after the
explosion of supernova SN1987A (which was first seen on February 23, 1987),
showed evidence for a bright source, separated from the SN by only 60 mas
(Nisenson et al, 1987).  This source (we will call it BS1) was seen at least
three times: by us 30 and 38 days after the explosion and by (Meikle et al,
1987) 50 days after the explosion. Two months later, (Chalabaev et al, 1989)
detected asymmetries in infrared speckle observations that were in
approximately the same north-south direction as BS1 which may have been
related to BS1. We observed SN1987A again 98 days after the explosion, but the
images failed to show conclusive evidence that BS1 was redetected.  While at
the time, there was much speculation as to plausible mechanisms for the
observations, see, for example (Rees, 1987; Piran and Nakamura, 1987; Colgate
et al, 1990), no fully satisfactory model was put forward to explain the
results. In this paper, we present a new reconstructed image that shows not
only a much cleaner reconstruction of BS1, but it also shows evidence of a
second bright spot on the opposite side of the SN. We will demonstrate that
this image is consistent with all the present data if we assume relativistic
jets emanating from the the SN. In addition, two groups have recently
suggested that at least some Gamma Ray Bursts (GRBs) could be explained as
highly collimated jets from supernovae (SNe) (Wang and Wheeler, 1998; Cen,
1998). A theoretical model for these jets, developed by Cen (Cen, 1999) to
explain GRBs is not only consistent with our original observations, but also
provides an explanation why BS1 faded by day 98.

\section{Observations}

Extensive observations of SN1987A were performed over the first two years
after the explosion, using the CfA PAPA detector and speckle interferometry
optics with the CTIO 4-meter telescope.  Details of observing procedures and
data reduction are in our earlier papers (Karovska et al, 1989; Papaliolios et
al, 1989; Karovska et al, 1991). These observations provided a unique high
angular resolution data base on the early history of SN1987A. The changing
size and shape of the expanding envelope was monitored, and substantial
asymmetries were detected.  Observations of the SN debris with the WF/PC
camera and the FOC over the last few years have confirmed the accuracy of our
measurements of the expansion velocity, as well as the asymmetries in the
debris shape and direction of the elongation (Pun, C.S.J. 1995; Pun, C.S.J
1997).  The direction of polarization ($200^{\circ}$) of the SN (Schwarz and
Mundt, 1987) was also found to be close to the direction of elongation of the
SN debris as measured by us ($200^{\circ}$), suggesting asymmetries in the
expanding debris.  However, our observations of BS1, that were seen at several
wavelengths, 30 and 38 days after the explosion, were viewed with some
skepticism, particularly since new observations made two months later did
not provide a certain redetection of BS1.

\section{Data Processing}

Recently, we went back to the original data and reprocessed them using newer,
more sensitive algorithms, to see if anything new could be learned about BS1
that would allow us to understand more about its nature.  The original
processing had used the Knox-Thompson (KT) algorithm (Knox and Thompson, 1974)
to recover both the modulus and phase of the image Fourier transform, in order
to obtain true image reconstructions. We used a large image magnification in
our speckle camera, so that the telescope diffraction limit would be well
oversampled. This resulted in a field of only 1.9 arseconds for the SN
observations, which was overfilled by the 2 to 3 arcseconds seeing disk.  For
the KT algorithm to work, the individual speckle frames must have a field size
that is at least twice the diameter of the seeing disk, so that adjacent
frequencies in the Fourier transform of each speckle frame are correlated in
phase.

In order to meet the KT requirement of field size and seeing disk size, each
speckle frame was multiplied by a gaussian mask which effectively narrowed the
"seeing disk" to less than one arcsecond.  This allowed the phase to converge
in the speckle average, but it also had the deleterious effect of reducing the
signal-to-noise ratio in the reconstructions, since all the speckles towards
the outside of the mask were discarded.

Between the time of our original processing of the data and the present, we
have implemented a new image reconstruction algorithm that does not require
masking the data with a gaussian.  Instead, we generate the speckle
autocorrelation using the unmasked data and then apply a modified version of
an iterative transform algorithm (ITA) (Fienup, 1984) to find the phase in the
Fourier transform and produce an image.  The major drawback in the ITA method
is that there is a $180^{\circ}$ ambiguity in the orientation of the
reconstruction.  However, the masked KT reconstructions clearly show BS1,
allowing a determination of the orientation and eliminating the ambiguity,
despite their poorer SNR.

The results of this processing on the SN data are consistent with our earlier
KT results but give cleaner reconstructions of BS1. In addition, the cleaner
images reveal a new feature that we believe could play an important role in
understanding not only BS1 but also the geometry of the whole SN system.
Figure 1 shows our new reconstruction from the data set taken with a 10 nm
wide filter centered on 654 nm (10 nm bandpass) from day 38.  On the left is a
false color image and on the right, a contour plot with plate scale and
orientation on it. The 654 nm data were far and away the best for SNR of all
our observations, since they were aquired with the smallest zenith angle, so
the atmospheric dispersion was small, the seeing was the best, and the signal
levels were the greatest (the peak sensitivity of the detector is in the red).
The plot shows the lowest 1\% contours with a linear display up to 10\%, and
then has contours at 20, 40, 60 and 80\%.  It very clearly shows BS1 separated
by four pixels from the SN, corresponding to $60 \pm 8$ mas angular
separation, with a magnitude difference from the SN of $2.7 \pm .2$
magnitudes.  It also reveals a second feature, in-line with BS1, on the
opposite side of the SN, separated by $160 \pm 8$ mas and somewhat fainter
than BS1 (a $4.2 \pm .2$ magnitude difference from the SN).  We refer to this
feature as BS2.  While BS2 is substantially fainter than BS1, it is well
above the noise floor in the new image (but not in the original KT
reconstructions). BS2 is $9.5 \sigma$ above the background where $\sigma$
is the standard deviation calculated from regions adjacent to the images.

There are a few additional point-like peaks in the image that are almost
certainly reconstruction noise. These peaks have a different character from
BS1 and BS2 since they do not show the roughly east-west elongation that is
seen in the SN, BS1 and BS2 features. This elongation is due to imperfectly
corrected atmospheric dispersion. The brightest of these other peaks
(north-east of the SN) is 0.7 magnitude fainter than BS2, and the other peaks
are more than 1 magnitude fainter (so they fall below the 1\% contour level in
the plot). These noise peaks are introduced when we deconvolve the SN image
with data recorded on a nearby bright, unresolved star, using a Wiener
filter. This deconvolution is critical in correcting the speckle
interferometry transfer function (Nisenson, 1989). Adjusting the Wiener filter
smooths the image and changes the noise characteristics.  For the image in
Figure 1, the filter was adjusted to cleanly separate BS1 from the SN. If the
filter is set so that the high frequencies in the image are reduced, the noise
peaks are eliminated.  BS1 and BS2 remain but have a broader shape and BS1
starts to merge with the SN image.

Three other data sets were processed in the same way: the images taken on day
38 with filters centered on 533 nm and 450 nm, and the image taken on day 30,
with the filter centered at 654 (all filters had a 10nm bandpass). The 533 nm
image from day 38 and the 654 nm image from day 30 image show the BS1 feature
more clearly than the KT reconstructions, but the reconstruction noise does
not allow a reliable identification of a feature corresponding to BS2. There
is a feature in the day 30 image in the location predicted if there were
linear motion from the date of the explosion. However, this image is noisier
than the 654 nm day 38 image, and the 2nd spot is at the same level
as the reconstruction noise.  The day 30 image also shows that the separation
of the centroid of BS1 and the SN appears to be about 1 pixel less than the
separation in the day 38 image, consistent with linear outward motion of the
feature for the 8 day time difference. The image taken with the 450 filter is
very noisy and no reliable identification of even BS1 could be made from
it. We also note that (Miekle et al, 1987) detected a bright feature 50 days
after the explosion, separated from the SN by 74 $\pm$ 8 mas in the same
direction as BS1. Assuming their detection and ours were the same phenomenon,
this measurement is also consistent with linear motion.

\section{Analysis}

If we interpret BS1 and BS2 as evidence for a jet and counter-jet, and we
assume that the two sources were ejected at the time of the
explosion, we can then calculate the geometry of the system. The distance to
SN1987A is nominally 50 kpc. At that distance, the angular separation of BS2
from the SN results in an apparent velocity that is superluminal
(V$_{apparent} = 1.24c$) where c is the velocity of light. This requires
that BS2's velocity is relativistic and that it appears to be superluminal
because it has a component of its velocity coming towards us
(blue-shifted). Since all relativistic velocities saturate at c, it is
reasonable to assume the actual velocities of BS1 and BS2 are equal. With that
assumption, the equation for the angle to the line of sight, $\theta$, is
given by (Pearson et al, 1981)

\begin{equation}
  \tan \theta = \frac{2 V_1 V_2}{(V_2 - V_1)}
\end{equation}

where V$_1$ and V$_2$ are the apparent velocities (normalized to c) of the
red-shifted and blue-shifted components, respectively. From the
apparent distances of BS1 and BS2 from the SN in Figure 1, the known time
between the explosion and the observations, and the distance to the SN, we can
immediately calculate $\theta$, the angle that the line joining BS1, the SN
and BS2 makes with the line-of-sight to the supernova.  We can then calculate
the actual velocity, $v/c$, of the components with the formula

\begin{equation}
  v/c = \frac{V_1} {\sin\theta_1 - V_1  \cos\theta_1}       
\end{equation}

In this case, V$_1$ = $.46 \pm .06$ and V$_2$ = $1.24 \pm .06$. Then, from
equation (1), $\theta$ is calculated to be $55^{\circ} \pm 3$ from the
line-of-sight, and, from equation (2), $v/c$ is .80.  The position angle for
BS1 is $194^{\circ} \pm 3$ and for BS2 is $14^{\circ} \pm 3$.

\section{Discussion}

This result is particularly interesting since HST (FOC) images of SN1987A
recorded in O {\sc III} (500.7 nm) show that the SN appears to be at the
center of a bipolar nebula in the shape of an hourglass with a bright ring at
its waist (Plait et al, 1995).  This ring appears to be centered on the SN and
is thought to be the boundary between the Red Supergiant wind from the earlier
stages of the SN progenitor evolution, and the faster wind from the Blue
Supergiant which was the last evolutionary stage before the explosion. The EUV
flash from the explosion excited the gas in the ring, and it has been slowly
dropping in brightness with time. The narrow elliptical shape probably
indicates that the outflow from the star had a preferential plane (likely to
be the SN equatorial plane) which is tipped by $44^{\circ}$ from the line of
sight if one assumes that the ring is circular.  This is not very different
from the angle ($55^{\circ}$) from the line-of sight that was calculated for
the axis along which BS1, the SN, and BS2 lie.  In addition, the position
angle on the sky of the minor axis of the ellipse is $179^{\circ} \pm 3$,
close to the position angle of BS1 and BS2 on the sky ($194^{\circ}$). Also,
the position angle of the jet is nearly aligned with the direction of
elongation of the SN debris.
 
This is highly suggestive that BS1 and BS2 are within 11 degrees from
perpendicularity to the equatorial plane of the SN if that plane is defined by
the observed ring. Since the blue-shifted spot (BS2) is to the north, this
would imply that southern part of the ring is closer to us.  However, if our
geometry is correct, then it conflicts with conclusions drawn from recent
measurements made with the Space Telescope Imaging Spectrograph (STIS)
(Michael et al, 1998). They see a blue-shifted hot spot on the northern part
of the ring, and a reverse shock from high-speed debris hitting the $H_2$
region inside the ring which is also blue-shifted to the north. Since the hot
spot and the higher density material are likely to be in the plane of the
ring, this implies that the northern part of the ring is nearer to us. The
geometry implied by the STIS measurements makes interpretation of the
phenomenon as a polar jet much less likely. As the debris starts to reach the
vicinity of the ring and starts to interact with it, these ambiguities should
be resolved.

Cen's model (Cen, 1999) predicts not only the color dependence and brightness
of BS1 and BS2, but also that their visual brightness would fade in the
interval between day 38 and day 98 after the explosion. His model also shows
that the brightness of the jet is proportional to its mass, so the brightness
ratio of BS1 and BS2 should be proportional to their mass ratio.  If the
intrinsic flux from each component was equal, the Lorentz factor would show
increased brightness ratio of the blue-shifted component compared to the
red-shifted component. Since the measured intensity of BS2 is approximately
four times fainter than BS1, the mass of BS1 must be much greater than the
mass of BS2 to compensate for the Lorentz effect. The spectral index of the
emission from the jet will also somewhat affect the brightness ratio and thus
the mass ratio. This difference in mass of the two components fits with the
model for a supernova jet (Cen, 1998). In his analysis, the jet is caused by
large asymmetries in supernova explosions, so one might expect unequal masses
in the jet and counter-jet.

We are reprocessing additional data, taken 98-100 days after the explosion,
with our new reconstruction algorithms to see if we can find a feature at the
position predicted by the velocity of the jet. By June 1, 1987, the SN had
increased in brightness by almost one magnitude, so even if BS1 had not faded,
as predicted, any certain detection will be very difficult because of the
expected large magnitude difference from the SN. Results of this reprocessing
will be published in a subsequent paper.

We thank Renyue Cen and John Raymond for many helpful discussions.

\newpage

\begin{figure}

\plotone{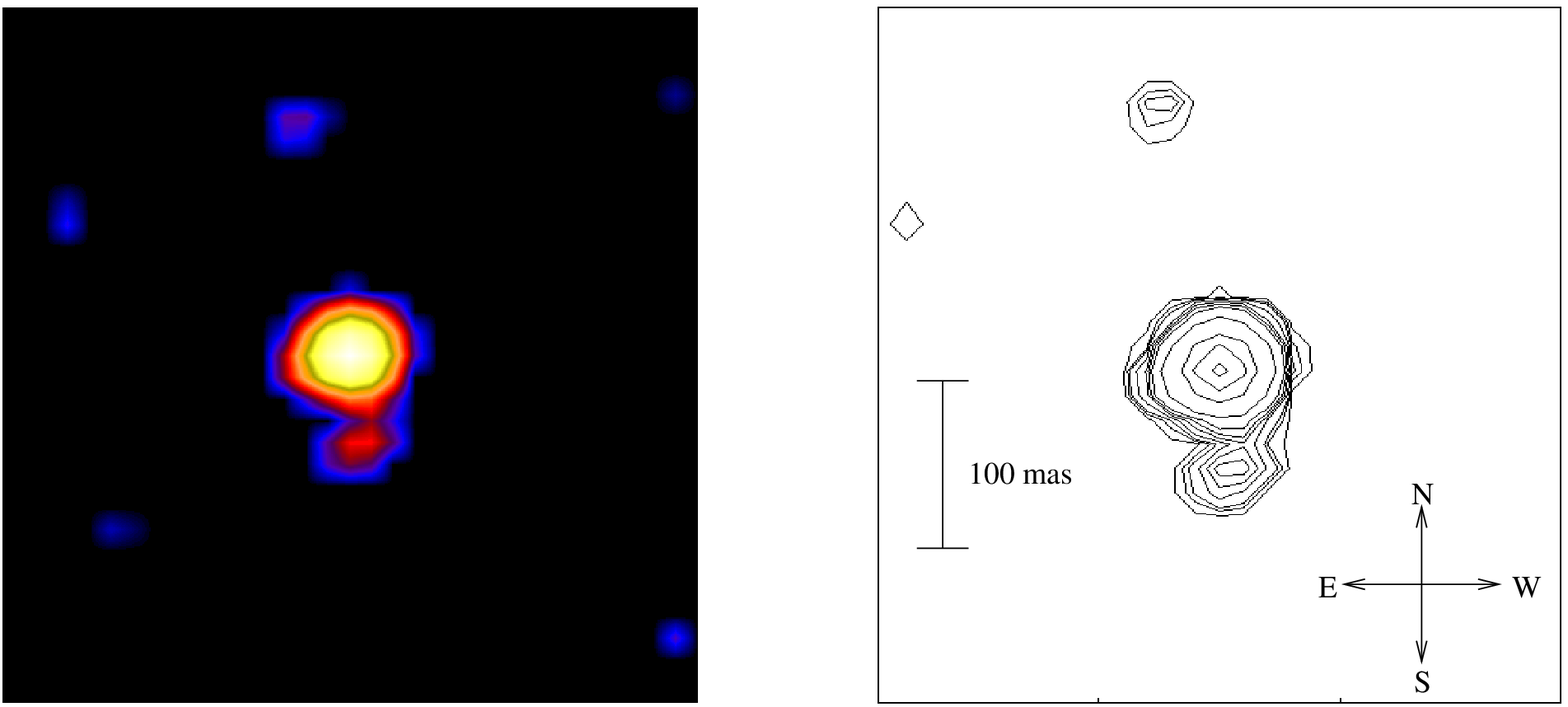}

\caption {New reconstructed image (left) and contour plot (right) of a 300
mas field around SN1987A from speckle data recorded with a 10 nm Wide
Filter centered at 653.6 nm. Note that two of the noise spikes (in the lower
left and right) do not show up in the contour plot because their intensity
falls below the lowest contour level (1\%).}

\end{figure}

\end{document}